\documentclass[aps,pre]{revtex4}

\def\la{\langle}
\def\ra{\rangle}
\usepackage[dvips]{graphicx}
\usepackage{epsfig} 
\usepackage[]{subfigure}
\usepackage[dvips]{color}

\usepackage{graphicx}
\usepackage{subfigure}
\usepackage{latexsym} 
\usepackage{amsmath,amscd}
\usepackage{times}

\font\gross=cmbx12 scaled \magstep0

\begin{document}
\baselineskip 0.45cm

\noindent
{\gross Dynamics of a stochastically driven Brownian particle \\
in one dimension}

\vspace*{1.0cm}\noindent
{\gross
S. L. Narasimhan$^1$ and A. Baumgaertner$^2$
}

\vspace*{0.5cm}\noindent
$^1$Solid State Physics Division, Bhabha Atomic Research Center, Mumbai - 400085, India\\
$^2$Institute of Solid State Research, Research Center, Juelich, Germany

\vspace*{1.0cm}\noindent

{\gross Abstract}
~We present a study on the dynamics of a system consisting of a pair of
hardcore particles diffusing with different rates.
We solved the
drift-diffusion equation for this model in the case when one particle,
labeled F,
drifts and diffuses slowly towards the second particle, labeled M.
The displacements of particle M exhibits a crossover from diffusion to
drift at a
characteristic time which depends on the rate constants.
We show that the positional fluctuation of M
exhibits an intermediate crossover regime of subdiffusion separating
initial and asymptotic diffusive behavior;
this is in agreement with the complete
set of Master Equations that describe the stochastic evolution of the
model.
The intermediate crossover regime can be considerably large depending
on the hopping probabilities of the two particles.
This is in contrast to the known crossover from diffusive to
subdiffusive behavior of a tagged particle that is in the interior of a
large single-file system on an unbound real line.
We discuss our model with respect to the biological phenomena of
membrane protrusions  where polymerizing actin filaments (F)
push the cell membrane (M).

\vspace*{0.5cm}\noindent
{\gross Keywords}~Brownian ratchet, driven dynamics, subdiffusion,  
non-equilibrium fluctuation


\section*{\leftline{\gross 1~~Introduction}}

Single-file diffusion of a system of hardcore particles is a well-studied 
process that provides a basic description of transport, for example, 
in fast ion transport through channels \cite{alberts,ab}, 
in zeolites \cite{kargar} and 
in superionic or organic conductors \cite{richards}. 
While their collective diffusion is like that of a set of independent 
particles, the diffusion of a {\it tagged} particle is known to be 
different - namely, its mean squared displacement (MSD) is proportional 
to $\sqrt{t}$ and its positions are Gaussian-distributed 
\cite{harris,jepsen,lebowitz,levitt,hahn,beijeren,kollman,percus}. 
It is an exactly solved problem in the case when the system consists of 
identical particles having the same diffusion constant. On the other hand, 
the case when their motion is characterized by a set of different diffusion 
constants is not readily amenable to an exact mathematical analysis. 
However, Ambj{\"o}rnsson {\it et al.}, \cite{ambjoern,lizana} have recently 
shown that the diffusion of a system consisting of just two hardcore 
particles with different diffusion constants on an unbounded 
(one dimensional) real line can be solved exactly; in particular, 
they have shown that the MSD of a tagged particle is proportional 
to $t$; they have also  presented Monte Carlo evidence to show that 
it crosses over from diffusive ($\propto t$) to 
subdiffusive ($\propto \sqrt{t}$) behavior only if it is in the interior 
of a large system. 

In this context, it is of interest to study the effect of constraining 
boundaries on the single-file diffusion of particles with different 
diffusion constants. Such boundary 
effects are relevant, for example,  to understand the physical mechanism 
underlying the process of cell protrusion where polymerizing actin filaments
push the cell membrane. 
The first physical description of the cell protrusion process was based 
on the Brownian Ratchet (BR) model \cite{peskin,mogilner}.
This is a one-dimensional 
two-particle representation for the filament-membrane system.
In this model,
the random diffusive motion of the cell membrane, represented as a Brownian obstacle,  
is rectified by the growing tip of a semi-rigid rod whose other end is fixed at the origin.  
Using the
stationary distribution for the gap between the tip of the flexible rod and the obstacle,
they could explain the experimentally observed load-velocity curves reasonably well.

In the two-particle model the membrane is represented by a single particle 
where the tension-induced
correlations among the constituents of a two-dimensional  flexible surface
are neglected.
In fact, a recent simulation study on two interacting random surfaces \cite{EW} representing 
the cell membrane and the actin cortex  has indicated the importance of
correlation \cite{Simha10}.
The representation of a long semi-rigid actin filament by a single particle is
reasonable as long as nucleation of new filaments near the leading edge of
the protruding membrane is negligible.
In this case,  a single actin filament polymerizes and 
depolymerizes at its both ends such that shrinking and growing of its length
would never lead to dissolvation. This assumption is justified based on the observation
that {\sl in vivo} the ATP-mediated 
(de)polymerization processes \cite{Pollard86}
can yield an effective `treadmilling' \cite{BrayBook} of a 
filament which, on average, depolymerizes only at its `minus' end and polymerizes only
at its `plus' end at the same rate such that its average length 
remains constant.
Therefore, the theoretical {\sl ansatz} to consider only the growing plus end,
which is located near the protruding edge of the cell membrane, 
and to model this by a single particle is reasonable.
Furthermore, in the two-particle model the assumption is made that the
whole filament is prohibited to perform large-scale thermal motion
(e.g. diffusion). {\it In vivo} this immobility is mostly achieved by
strong adhesive bonds connecting the filament via the cell membrane to the underlying 
substrate on which the cell crawls. 
Therefore, in the present two-particle model the immobility of the filament is presumed,
and 
the only mechanism by which a cell can
be moved is by pushing the leading front of the cell membrane (particle M)
by the polymerizing plus end of a filament (drifting particle F).

We present an exact analysis of a two-particle system 
drift-diffusing 
on the positive real line in the case when particle F drifts and diffuses slowly towards 
particle M.
The fluctuation in the position of the membrane particle M exhibits 
diffusive behavior at short, as well as at long times with
an intermediate crossover-regime that is due to the onset of hardcore interaction between 
the particles. We show that this is in agreement with the results obtained from the set of
Master equations that describe the stochastic evolution of the two-particle model. On the other 
hand, fluctuation in the position of the pushing 'filament' particle F is diffusive without 
any such crossover behavior.  
The mean displacement of the Brownian particle M 
('membrane') crosses over from an initial 
diffusive behavior to an asymptotic drift behavior. 
For the sake of completeness we consider also the case of a repulsive
boundary for F at the origin.

In the next section, we describe the one dimensional drift-diffusive motion of a pair of 
hardcore particles in terms of a set of hopping rates on a regular lattice and present a 
discussion
of this process in the continuum limit. In section III, we compare 
the results obtained by solving the two-particle drift-diffusion equation with those
obtained by numerically integrating the set of master equations, given in the Appendix, that 
describes the hopping
process on a lattice.

\bigskip

\section*{\leftline{\gross 2~~Two hardcore particles on a 1d lattice:} 
\\ {\gross Drift-Diffusion equation for the joint probability distribution}}  

\bigskip

\begin{figure}[hbt] 
  \begin{center}
    \includegraphics [width=0.5\textwidth ]{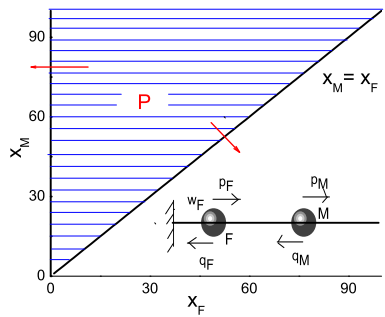}   
    \caption{A schematic of a system of two hardcore particles in the presence of a reflecting boundary. In the discrete-time picture,  
             the 'driven' particle, M, jumps right (or left) with {\it a priori} probability $p_M$ (or $q_M = 1-p_M$) at every
             instant of time. The 'driving' particle, F, jumps right (or left) with {\it a priori} probability   
             $p_F$ (or $q_F = 1-p_F$) {\it once in} $1/w_F$ {\it instants of time} on the average. Shaded region denotes the domain 
             of two-particle drift-diffusion.   }
    \label{fig:Model}
  \end{center}
\end{figure}

Let a pair of particles, labeled F and M, be at positions $n_F$ and $n_M$ 
respectively on a one dimensional lattice; hardcore interaction ensures 
that $n_F < n_M$ at all times, if that was the case at $t=0$. Equivalently, 
the distance of separation between these particles, $n \equiv (n_M - n_F)$, 
satisfies the inequality  $n \geq 1$ at all times. 
A schematic illustration of the model is shown in Fig.\ref{fig:Model}.

Consider a discrete time process. At every instant of time, particle M moves 
one step to the right or to the left with {\it a priori} probability $p_M$ 
or $(1-p_M)$ respectively; on the other hand, particle F may either move 
or stay put. 
Let $w_F$ be the probability that F will move at any given 
instant of time. Then, on the average, F moves just once during the period 
of time in which M has moved $1/w_F$ times. It is clear that $w_F$ also 
denotes the {\it a priori} probability that both F and M will move 
simultaneously at a given instant of time. 
Given that F moves, let it move one step to the right or to the left with 
{\it a priori} probability $p_F$ or $q_F \equiv (1-p_F)$ respectively. 
Of course, with respect to the physico-biological situation,
 as described in the Introduction, the practically reasonable case 
 is $q_F < p_F$, which implies $n_F > 0$.

So, when both 
the particles jump simultaneously in the same direction, the separation 
distance, $n$, does not change; when they jump simultaneously in opposite 
directions, $n$ changes by $\pm 2$; on the other hand, when particle F does 
not jump, $n$ changes by $\pm 1$. Let $q_0^{\pm}, q_1^{\pm}, q_2^{\pm}$ denote 
the probabilities per unit time that the change in separation distance 
$\delta n = 0, \pm 1, \pm 2$ respectively. They are given by
\begin{eqnarray}
\label{eq:q012}
q_0^{+} & = & w_F p_F p_M \nonumber \\
q_0^{-} & = & w_F (1-p_F)(1-p_M) \nonumber \\
q_1^{+} & = & (1-w_F)p_M \nonumber \\
q_1^{-} & = & (1-w_F)(1-p_M) \\
q_2^{+} & = & w_F(1-p_F)p_M \nonumber \\
q_2^{-} & = & w_Fp_F(1-p_M) \nonumber
\end{eqnarray}
and they all add up to unity. 

The physically motivated constraint that the particles can only move on the positive real line, $n_M > n_F \geq 1$ (on a lattice), as well as the probability, $w_F$, for them to move simultaneously imply that we have different sets of Master Equations for the joint probability, $P(n_F,n_M;t)$ corresponding to the cases $n = 1, 2, \geq 3$ respectively (see, Appendix).

In the continuum description, the joint probability for the positions of the hardcore particles, $x_F$ and $x_M$, satisfies the following drift-diffusion equation,
\begin{eqnarray}
\frac{\partial P(x_F,x_M;t)}{\partial t} & = & \left( D_F\frac{\partial ^2 P(x_F,x_M;t)}{\partial x_F ^2} - 
                                              \mu _F \frac{\partial P(x_F,x_M;t)}{\partial x_F}\right) + \nonumber \\ 
                                         &   & \left( D_M\frac{\partial ^2 P(x_F,x_M;t)}{\partial x_M ^2} - 
                                              \mu _M \frac{\partial P(x_F,x_M;t)}{\partial x_M}\right)
\label{eq:ddeq1}
\end{eqnarray}
which is, in fact, the continuum version of the unrestricted Master Equation, Eq(~\ref{eq:neq3}). Hardcore interaction between the particles constrained to be on the positive real line implies that $x_M > x_F \geq 0$ at all times. In our model, particle M is purely diffusive whereas particle F can drift as well. Therefore, as shown in Appendix (f), the corresponding diffusion and drift coefficients are given by,
\begin{eqnarray}
\label{eq:dmudefns}
D_F & = & \frac{1}{2}(q_2^{+} + q_2^{-} + q_0^{+} + q_0^{-}) \nonumber \\ 
    & = & \left(\frac{1}{2}w_F(1-w_F) + 2w_F^2p_F(1-p_F)\right) \ \approx \frac{1}{2}w_F \quad \mbox{($w_F$ small)}
                                                                                                \nonumber \\
\mu _F & = & - q_2^{+} + q_2^{-} + q_0^{+} - q_0^{-}  =  w_F(2p_F - 1) \nonumber \\ \\
D_M & = & \frac{1}{2}(q_2^{+} + q_2^{-} + q_1^{+} + q_1^{-} + q_0^{+} + q_0^{-})  =  \frac{1}{2} \nonumber \\
\mu _M & = & q_2^{+} - q_2^{-} + q_1^{+} - q_1^{-} + q_0^{+} - q_0^{-}  =  0 \nonumber
\end{eqnarray}
The initial condition for this problem could be specified by the joint probability distribution,
\begin{equation}
P(x_F,x_M;t=0) = \delta (x_F - x_F^0)\delta (x_M - x_M^0)
\label{init1}
\end{equation}
where $\delta (x)$ is the Dirac delta function. The first boundary condition,
\begin{equation}
\left( D_M\frac{\partial P(x_F,x_M;t)}{\partial x_M}-D_F\frac{\partial P(x_F,x_M;t)}{\partial x_F}\right)_{x_F=x_M}
= \left[ (\mu _M - \mu _F)P(x_F,x_M;t)\right]_{x_F=x_M}
\label{eq:bound1}
\end{equation}
expresses the fact that the particles cannot pass each other ($x_F < x_M$). The second boundary condition,
\begin{equation}
\left( D_F\frac{\partial P(x_F,x_M;t)}{\partial x_F} - \mu _F P(x_F,x_M;t)\right)_{x_F=0} = 0
\label{eq:bound2}
\end{equation}
ensures that there is no current across the boundary at $x_F=0$. These two boundaries define a wedge as the domain for the joint distribution $P(x_F,x_M;t)$. The question is whether a separable solution to Eq.(\ref{eq:ddeq1}) can be found. We first transform the variables, $\{x_F,x_M\}$, into a pair of 'collective' variables, say $\{x,r\}$.

Inter-particle separation, $r \equiv x_M-x_F$, could be fixed as one of the new variables. For the other variable, we set $x= c_Mx_M + c_Fx_F$ and try to fix the dimensionless constants, $c_M$ and $c_F$, by requiring that the transformed drift-diffusion equation also has the same form as Eq.(\ref{eq:ddeq1}). This requirement leads to the condition that the coefficient of ${\partial ^2P(x,r;t)}/{\partial x}{\partial r}$ vanishes:
\begin{equation}
D_Mc_M - D_Fc_F = 0,
\end{equation}
This implies that
\begin{equation}
c_M\sqrt{\frac{D_M}{D_F}} = c_F\sqrt{\frac{D_F}{D_M}} = c
\end{equation}
where $c$ is an arbitrary constant. We set $c=1/2$ so that $x$ will be the position of the center-of-mass of the system when the particles are of the same mass and have the same value for their diffusion constants ($D_M=D_F$). In general, $x$ is not the center-of-mass coordinate because the equality, $c_M+c_F = 1$, will be satisfied only when the particles have the same mass and also when $D_M=D_F$. Nevertheless, with the choice $c=1/2$, we have the transformation, 
\begin{eqnarray}
\label{eq:xandr}
x & = & \frac{1}{2}\left(\sqrt{\frac{D_M}{D_F}}\ x_F + \sqrt{\frac{D_F}{D_M}}\ x_M\right) \nonumber \\
 \\
r & = & x_M - x_F \nonumber
\end{eqnarray}
under which Eq.(\ref{eq:ddeq1}) becomes
\begin{equation}
\frac{\partial P(x,r;t)}{\partial t} = \left( D_x\frac{\partial ^2 P(x,r;t)}{\partial x^2} - 
                                              \mu _x \frac{\partial P(x,r;t)}{\partial x}\right) +  
                                       \left( D_r\frac{\partial ^2 P(x,r;t)}{\partial r^2} + 
                                              \mu _r \frac{\partial P(x,r;t)}{\partial r}\right)
\label{eq:ddeq2}
\end{equation}
with the initial condition,
\begin{equation}
P(x,r;t=0) = \left( \frac{D_F+D_M}{2\sqrt{D_FD_M}}\right)\ \delta (x-x^0)\ \delta(r-r^0)
\label{eq:init2}
\end{equation}
where
\begin{eqnarray}
\label{eq:defn}
D_x & = & (D_M+D_F)/4 \nonumber \\
D_r & = & D_M+D_F \nonumber \\
\mu _x & = & \frac{1}{2}\left(\sqrt{\frac{D_M}{D_F}}\ \mu _F + \sqrt{\frac{D_F}{D_M}}\ \mu _M \right) \\
\mu _r & = & \mu _F - \mu _M \nonumber \\
x^0 & = & \frac{1}{2}\left(\sqrt{\frac{D_M}{D_F}}\ x_F^0 + \sqrt{\frac{D_F}{D_M}}\ x_M^0 \right) \nonumber \\
r^0 & = & x_M^0 - x_F^0 \nonumber
\end{eqnarray}
The boundary condition, Eq.(\ref{eq:bound1}), that ensures hardcore repulsion between the particles now transforms into the following condition,
\begin{equation}
\left[ D_r\frac{\partial P}{\partial r}+ \mu _r P \right]_{r=0} = 0
\label{eq:bound3}
\end{equation}
which is the reflecting boundary condition at $r=0$ ({\it i.e.,} $x_F=x_M$) for the above drift-diffusion equation for $P(x,r;t)$. The other boundary condition, Eq.(\ref{eq:bound2}), transforms into
\begin{equation}
\left[ \frac{1}{2}\sqrt{D_MD_F}\ \frac{\partial P}{\partial x} - 
                              D_F\frac{\partial P}{\partial r} - 
                              \mu _F P\right]_{x = \frac{1}{2}\sqrt{\frac{D_F}{D_M}}\ r} = 0
\label{eq:bound4}
\end{equation} 
It is clear that the wedge in $\{x_F,x_M\}$-space has transformed into a wedge in the $\{x,r\}$-space. Moreover, the above boundary condition implies that separable solution to Eq.(\ref{eq:ddeq2}) is not possible in the transformed space.

\bigskip

\noindent{{\bf A. No reflecting boundary at $x_F=0$}:

\bigskip
\begin{figure}[hbt] 
  \begin{center}
    \includegraphics [width=0.5\textwidth ]{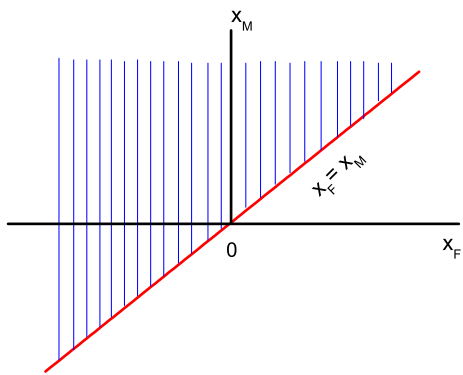}   
    \caption{Shaded area is the domain of the joint probability distribution, $P(x_F,x_M;t)$, with 
    only one reflecting boundary  
             corresponding to the hard-core repulsion between the particles.}
    \label{fig:Model1}
  \end{center}
\end{figure}

One way to ensure separable solution to Eq.(\ref{eq:ddeq2}) is to ignore the 
boundary condition, Eq.(\ref{eq:bound2}), and hence Eq.(\ref{eq:bound4}). This amounts 
to assuming that the growing filament does not 
degrade to a monomer in any arbitrary time span of interest; so, particle 
F moves freely as long as it is to the left of particle M. 
In this case, we may now try a product solution to Eq.(\ref{eq:ddeq2}) 
of the form,
\begin{equation}
P(x,r;t) = \left( \frac{D_F+D_M}{2\sqrt{D_FD_M}}\right) P(x;t) P(r;t)
\label{eq:cmdist5}
\end{equation}
so that $P(x;t)$ and $P(r;t)$ satisfy the equations:
\begin{eqnarray}
\label{eq:ddeq3}
\frac{\partial P(x;t)}{\partial t} & = & D_x \frac{\partial ^2P(x;t)}{\partial x^2} - 
                                         \mu _x \frac{\partial P(x;t)}{\partial x} \nonumber \\
\\ 
\frac{\partial P(r;t)}{\partial t} & = & D_r \frac{\partial ^2P(r;t)}{\partial r^2} + 
                                         \mu _r \frac{\partial P(r;t)}{\partial r} \nonumber
\end{eqnarray}
The first one is a free-boundary equation with the initial condition, $P(x;t=0) = \delta (x-x^0)$, 
whereas the second one, with the initial condition $P(r;t=0) = \delta (r-r^0)$, is subject to the 
boundary condition given by Eq.(\ref{eq:bound3}). Solutions can be written down immediately 
\cite{chandra}: 
\begin{eqnarray}
\label{eq:cmdist6} 
P(x;t) & = & g\left( \frac{x-x^0-\mu _xt}{\sqrt{4D_xt}}\right) \nonumber  \\
\\
P(r;t) & = & g\left( \frac{r-r^0+\mu _rt}{\sqrt{4D_rt}}\right) + 
             e^{\mu _r r^0/D_r}g\left( \frac{r+r^0+\mu _rt}{\sqrt{4D_rt}}\right) + \nonumber \\
       &   &\qquad \qquad \qquad
        \frac{\mu _r}{D_r}e^{-\mu _r r/D_r}\int _{y=r^0}^{\infty}g\left( \frac{y+r-\mu _rt}{\sqrt{4D_rt}}\right)dy
                                                                                             \nonumber
\end{eqnarray}
where $g(\xi /\sigma)$ is the normalized gaussian function given by
\begin{equation}
g(\xi /\sigma) = \frac{1}{\sqrt{\pi \sigma ^2}}\ e^{-\xi ^2/\sigma ^2}
\label{eq:gaussian}
\end{equation}
The distribution $P(r;t)$ will become asymptotically stationary when $\mu _r > 0$; this implies the 
condition $w_F(2p_F-1) > (2p_M-1)$ which is to say that the effective rightward drift of particle F 
should be more than that of particle M. It is clear from the definition, Eq.(\ref{eq:xandr}), that 
stationary value for $\la r \ra$ does not imply stationary values for $\la x_M\ra$ and $\la x_F\ra$.

\bigskip

\noindent{\bf B. Reflecting boundary at $x=0$}:

\bigskip
\begin{figure}[hbt] 
  \begin{center}
    \includegraphics [width=0.5\textwidth ]{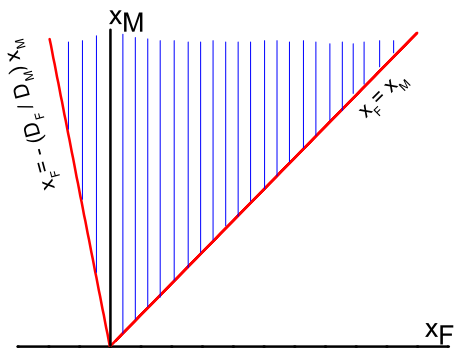}   
    \caption{Shaded area is the domain of the joint probability distribution, $P(x_F,x_M;t)$, with one reflecting boundary  
             at $x=0$ and the other corresponding to the hard-core repulsion between the particles.}
    \label{fig:Model2}
  \end{center}
\end{figure}

In the Brownian Ratchet model \cite{peskin, mogilner} a reflecting boundary at the origin is introduced
in order to provide a load which prohibits backflow of the actin filament.
In this model the (de)polymerizing filament is represented by a rod where the plus end can
(de)polymerize and the 
minus end of the filament is fixed at the origin.
Since particle F represents the plus end of the rod, which has to be of 
nonzero length, we have a reflecting boundary for particle F at the origin.

This
reflecting boundary at $x_F=0$ transforms into the boundary condition 
Eq.(\ref{eq:bound4}), which rules out a separable form for $P(x,r;t)$. On the other hand, 
a separable solution to Eq.(\ref{eq:ddeq2}) can still be obtained if we arbitrarily impose the 
following boundary condition at $x=0$: 
\begin{equation}
\left[ D_x\frac{\partial P(x;t)}{\partial x}- \mu _x P(x;t) \right]_{x=0} = 0
\label{eq:bound5}
\end{equation}
From the definition, Eq.(\ref{eq:xandr}), we see that $x=0$ implies $x_F=-(D_F/D_M)x_M$; hence the domain for the joint distribution, $P(x_F,x_M;t)$, is the wedge schematically shown in Fig.\ref{fig:Model2}. It is unphysical if we insist that particle F represents the tip of a polymer rod; yet, for small $w_F$ (by definition, Eq.(\ref{eq:dmudefns}), $(D_F/D_M)\approx w_F$), we may expect to have an approximate solution to the model.

Assuming separable form for $P(x,r;t)$ (Eq.(\ref{eq:cmdist5})), the above boundary condition at $x=0$ leads to the solution, \cite{chandra}
\begin{eqnarray}
P(x;t) & = & g\left( \frac{x-x^0-\mu _xt}{\sqrt{4D_xt}}\right) + 
             e^{-\mu _x x^0/D_x}g\left( \frac{x-x^0-\mu _xt}{\sqrt{4D_xt}}\right) - \nonumber \\
       &   &\qquad \qquad \qquad
        \frac{\mu _x}{D_x}e^{\mu _x x/D_x}\int _{y=x^0}^{\infty}g\left( \frac{y+x+\mu _xt}{\sqrt{4D_xt}}\right)dy 
\label{eq:cmdist7}
\end{eqnarray}
The distributions, $P(x;t)$, will become asymptotically stationary when $\mu _x<0$. From the definitions, Eqs.(\ref{eq:dmudefns},\ref{eq:defn}), we see that $\mu _x = w_F(p_M+p_F-1)$ and so $\mu _x < 0$ implies the condition $p_F < q_M$. This will be realized when the leftward drift of particle M is more than the rightward drift of particle F. Interestingly, stationary value for $\la x \ra$ implies, by definition Eq.(\ref{eq:xandr}), stationary values for $\la x_M\ra$ and $\la x_F\ra$ also, which in turn implies stationary value for $\la r\ra$. Hence, the condition $\mu _x < 0$ is enough to ensure that the joint probability distribution, $P(x,r;t)$, becomes stationary. 

\bigskip

\noindent{\bf C. Tagged particle distributions}:

\bigskip
 
Transforming back to the variables, $x_F$ and $x_M$, with appropriate Jacobian prefactor, the above solutions for $P(x;t)$ and $P(r;t)$ lead to the joint distribution $P(x_F,x_M;t)$ and hence to the tagged particle distributions:
\begin{eqnarray}
\label{eq:tagdist2}
P(x_F;t) & = & \int _{x_F^+}^{\infty} dx_M P(x_F,x_M;t) \nonumber \\
\\
P(x_M;t) & = & \int _{-\infty}^{x_M^-} dx_F P(x_F,x_M;t) \nonumber
\end{eqnarray} 
The upper limit $x_M^-$ suggests that $x_F$ may be arbitrarily close but never equal to $x_M$; similarly, the lower limit $x_F^+$ suggests that $x_M$ may be arbitrarily close but never equal to $x_F$. The Mean Squared Displacement (MSD) of a tagged particle, 
$\sigma _{M,F}^2$, may then be obtained from its corresponding distribution.

\bigskip

\section*{\leftline{\gross 3~~Results and Discussions}}

\bigskip

\noindent{\bf A. Inter-particle distance, $r$}:

\bigskip

Stationarity for the distribution $P(r;t)$ in Eq.(\ref{eq:cmdist6}) is ensured by the condition $\mu _r > 0$ which, from Eqs.(\ref{eq:dmudefns},\ref{eq:defn}), implies $w_F(2p_F-1) > (2p_M-1)$. In this case, the asymptotic stationary form of $P(r)$ is given by
\begin{equation}
\label{eq:StationaryPr}
P(r) = P(r;t\to \infty) = \frac{\mu _r}{D_r}\ e^{-\mu _r r/D_r}
\end{equation}
which leads to the asymptotically stationary values for the moments:
\begin{eqnarray}
\label{eq:asymptr}
\la r\ra _{t\to \infty} & = & \frac{D_r}{\mu _r} \nonumber \\
\\
\la r^2 \ra _{t\to \infty} & = & 2\left( \frac{D_r}{\mu _r}\right)^2 \nonumber 
\end{eqnarray}
Initial time-dependence of these moments and their approach to these stationary values can be obtained from their full forms:
\begin{eqnarray}
\label{eq:MomentPr1}
\la r\ra & = & \frac{D_r}{2\mu _r}\mbox{erfc}(r^{0-}) + \frac{1}{2}(r^0-\mu_rt)\mbox{erfc}(-r^{0-}) - \nonumber \\ 
&   & \frac{D_r}{2\mu _r}\ e^{\mu _rr^0/D_r}\mbox{erfc}(r^{0+}) + \sqrt{\frac{D_rt}{\pi}}e^{-(r^{0-})^2}    
\end{eqnarray}
\begin{eqnarray}
\label{eq:MomentPr2}                                           
\la r^2 \ra   & = & \left( \frac{D_r}{\mu _r}\right) ^2
                    \left[(\mbox{erfc}(r^{0-}) - e^{\mu _rr^0/D_r}\mbox{erfc}(r^{0+})\right]- \nonumber \\
              &   & \frac{2D_r}{\mu _r}\left[ \sqrt{\frac{D_rt}{\pi}}e^{-(r^{0-})^2} -  
                    \frac{1}{2}(r^0+\mu_rt)\mbox{erfc}(r^{0+})\right]+ \nonumber \\                    
              &   & \frac{1}{2}\left[(\mu _rt)^2 + 2(D_r-\mu _rr^0)t + (r^0)^2\right]\mbox{erfc}(-r^{0-}) +
                    (r^0-\mu _rt)\sqrt{\frac{D_rt}{\pi}}e^{-(r^{0-)^2}}
\end{eqnarray}
where we have the definitions,
\begin{eqnarray}
\label{eq:erfcDefn}
r^{0+} & = & \frac{r^0+\mu _rt}{\sqrt{4D_rt}} \nonumber \\
r^{0-} & = & \frac{r^0-\mu _rt}{\sqrt{4D_rt}} \\
\mbox{erfc}(x) & = & 1-\frac{2}{\sqrt{\pi}}\int _0^x e^{-y^2}dy \nonumber
\end{eqnarray}
The 'early' time behavior ($t < r^0/\mu _r$) of the moments can be guessed by recognizing that $r^{0\pm} \sim r^0/\sqrt{4D_rt}$ is large and therefore, in the limiting case $\mu _r \to 0$ for example, the average separation $\la r \ra$ increases proportional to $\sqrt{t}$; larger the value of $\mu _r$, shorter will be the growth regime $\la r \ra \propto \sqrt{t}$ because it will attain its stationary value $D_r/\mu _r$ faster.

It will be interesting to compare these results with those obtained by numerically integrating the Master equations, Eqs.
\ref{eq:sepdist}), given in the Appendix (e). We have presented in Fig.\ref{fig:SepDist}(a) the distributions of separation-distance obtained by numerically integrating these equations for 133 and 3398 time steps respectively; the distribution is clearly exponential at $t = 3398$ - namely, $P(r) \sim 0.1e^{-0.096r}$ suggesting that the average value $\la r \ra \sim 10.42$. For the parameters used in the computation, the stationary value of $\la r \ra$ should be $D_r/\mu _r = (1+w_F)/2w_F = 10.5$ according to Eq.(\ref{eq:asymptr}). Even the normalizing constant ($\sim 0.1$) obtained from the Master equations is close to the expected value of $\mu _r/D_r \sim 0.095$.

In fact, the diffusion coefficient, $D_n$, obtained from Eq.(\ref{eq:sepdist}) in the continuum limit for the case $n\geq 3$ is equal to $D_r$ (Eq.(\ref{eq:defn}) for $p_M = 1/2$ and $w_F$ small; but, $\mu _n = \mu _r$ whatever be the value of $p_M$.

The small difference in the approach of $\la r \ra$ to its stationary value, observed in Fig.\ref{fig:SepDist}(b) for the case $p_M=1/2$, implies that the distribution obtained by integrating the Master Equations is not completely stationary even after 3398 time steps for the lattice size chosen; moreover, finite size effects could make slight differences because of the need to have boundary equations for $n = L$ and $ n = (L-1)$ similar to what we have for $n = 1$ and $n = 2$. 

\bigskip

\noindent {\bf B. Two-particle variable, $x$}:

\bigskip

In the case when there is no reflecting boundary at $x_F=0$, the variable $x$ is unbounded and hence its  distribution $P(x;t)$ is a gaussian (Eq.(\ref{eq:cmdist6})) with the following moments:
\begin{eqnarray}
\label{freexmoms}
\la x \ra & = & x_0 + \mu _xt \nonumber \\
\\
\la x^2 \ra & = & 2D_xt + (x_0+\mu _xt)^2 \nonumber
\end{eqnarray}
On the other hand, when the variable $x$ is constrained to be positive, its distribution $P(x;t)$ is given by Eq.(\ref{eq:cmdist7}). The corresponding moments can be obtained from Eqs.(\ref{eq:MomentPr1}) and (\ref{eq:MomentPr2}) by making the following replacements:
\begin{eqnarray}
\label{eq:rtox}
r \leftrightarrow x & ; & \qquad r^0 \leftrightarrow x^0 \nonumber \\
\mu _r \leftrightarrow -\mu _x & ; & \qquad D_r \leftrightarrow D_x \\
r^{0-} \leftrightarrow x^{0+} & ; & \qquad r^{0+} \leftrightarrow x^{0-} \nonumber
\end{eqnarray}
The moments have the following asymptotic behavior ($\mu _x > 0$):
\begin{eqnarray}
\label{eq:asymptx}
\la x \ra _{t \to \infty} & = & \mu _xt + \left( x^0 + \frac{D_x}{\mu _x}\ e^{-\mu _xx^0/D_x}\right) \nonumber \\
\\                                                                                                                 
\la x^2 \ra _{t\to \infty} & = & (\mu _xt)^2 + 2(D_x + \mu _xx^0 + D_x e^{-\mu _xx^0/D_x})t + \mbox{const.} \nonumber
\end{eqnarray}

\bigskip

\noindent {\bf C. Mean Squared Displacement of a tagged particle}:

\bigskip

It is tempting to see how the mean-squared fluctuation in the position of a tagged particle, say M, would be related to those corresponding to the variables, $r$ and $x$, given the fact that the transformation $(x_M,x_F) \longleftrightarrow (r,x)$ is one-to-one. Quite surprisingly, the following ansatz,  
\begin{equation}
\label{eq:sigmaMF}
\sigma _M^2 \equiv  \la x_M^2\ra - \la x_M\ra ^2 = \frac{4 D_M D_F}{(D_M+D_F)^2}\left(
                                             \sigma _x^2 + \frac{1}{4}\frac{D_M}{D_F}\ \sigma _r^2 \right)
\end{equation}
with $\sigma _{x,\ r}^2$ calculated using the distributions, $P(x;t)$ and $P(r;t)$ given by Eq.(\ref{eq:cmdist7}) and Eq.(\ref{eq:cmdist6}) respectively, leads to a striking agreement with those obtained from the Master equations.

In Fig.\ref{fig:AvPosSigmaM}(a), we have presented $\sigma _M$ obtained by using the above ansatz for the case when we have a reflecting boundary at $x=0$ (continuous line A) as well as from the Master Equations (open circles). The parameters are $w_F = 0.05; p_F = 1; p_M = 1/2$ and the agreement is quite good. In the same figure, continuous line B represents $\sigma _M$ obtained using the above ansatz but for the case when there is no reflecting boundary at $x_F=0$. Its long-time deviation from both the case corresponding to a reflecting boundary at $x=0$ (continuous line A) and the Master eqns.(open circles) indicates that it does not represent the basic phenomenology of the Master eqns. 

On the other hand, a similar ansatz for the average positions
\begin{equation}
\label{eq:AvPosMF}
\la x_M\ra = \frac{2\sqrt{D_M D_F}}{D_M+D_F}\left(
    \la x \ra + \frac{1}{2}\sqrt{\frac{D_M}{D_F}}\ \la r \ra \right)
\end{equation}
does not lead to such a good agreement with what we obtain from the Master 
Equations at early times. However, it leads to qualitatively similar 
asymptotic behaviors - namely, the drift velocities are proportional to 
each other, the proportionality constant being $w_F$-dependent. For example,
in (Fig.\ref{fig:AvPosSigmaM}(b)), the average 
velocity of particle M obtained from the Master Equations (open circles) is 
presented along with that obtained using the ansatz Eq.(\ref{eq:AvPosMF}) 
(line M). It is quite clear that their asymptotic drift velocities are just 
proportional to each other even though their early time behaviors are 
different. Line F in the figure is the velocity of particle F obtained by 
using the above ansatz. While it takes some time, say $\tau _M^{d}$ that 
depends on $w_F$ and $p_F$, for M to start 
drifting with constant velocity, F starts drifting almost from the 
beginning.  

In Fig.\ref{fig:AvPosSigmaM1}(a), we have presented $\sigma _M$ obtained by using the above ansatz for the case when we have a reflecting boundary at $x=0$ but with $w_F = 0.1$ (continuous line); the agreement with the Master Equations' data (open circles) is reasonably good. On the other hand, in the case $w_F = 0.2$ presented in Fig.\ref{fig:AvPosSigmaM1}(b), there is no agreement between the analytically computed data (continuous line) and those obtained from the Master equations(open circles). This demonstrates, as mentioned earlier, that results obtained with an unphysical reflecting boundary at $x=0$ may agree with those obtained with the physically meaningful reflecting boundary at $x_F=0$ for small values of $w_F$ (namely, $w_F \leq 0.1$ with $p_F = 1$ and $p_M = 1/2$). 

In Fig.\ref{fig:Sigma}(a), we have shown how the crossover-behavior 
of $\sigma _M$ contrasts with the monotonously increasing behavior 
of $\sigma _F$. It is only when particle F has 
no effective drift towards particle M that such a crossover is non-existent 
(see Fig.\ref{fig:Sigma}(b)); larger the effective drift, sooner is the 
crossover seen. It remains to be answered whether it is due to the 
reflecting boundary at the origin.

We have shown in Fig.\ref{fig:SigmaRBpF09} the fluctuation in the position of M for the cases, 
with as well as without a reflecting boundary at $x=0$. The parameters used are mentioned 
in the caption. It is clear that the crossover becomes more pronounced and shifted to later 
times in the presence of a reflecting boundary than in its absence. In fact, the crossover is 
due to the smaller diffusivity of particle F than that of particle M ($D_F < D_M$). The effective 
drift of particle F leads to the saturation of the average separation-distance.

\section*{\leftline{\gross 4~~Summary and Conclusions}}

It is appropriate, at this juncture, to compare this model with the standard 
BR model \cite{peskin,mogilner}.
The ratchet mechanism, in the BR model, is due to 
a monomer squeezing itself in the gap between the filament-tip and the barrier
particle ({\it i.e}., the membrane). The resulting 
growth of the filament-rod is impeded by the inward (forced) drift of the barrier particle.
This leads to a steady state in which the average gap betwen the
filament tip and the barrier remains constant. The average ratchet velocity, in the 
steady state, is proportional to the net polymerization rate of the filament and is given by
\begin{equation}
\label{eq:BRvelocity}
v_{BR} = \delta \left( \alpha \ \mbox{exp}\left[ -\frac{Df}{k_BT}\right] - \beta \right)   
\end{equation}
where $\alpha$ and $\beta$ are the polymerization and depolymerization rates
respectively; $D$ is the diffusion constant of the barrier particle, which drifts
towards the filament-tip under the influence of the dimensionless force, $f/k_BT$; 
and, $\delta$ is the size of the intercalating monomer.

In our model, the BR-parameters ($\delta$, $\alpha$, $\beta$, $D$ and $f/k_BT$) are
all lumped into the parameters $w_F$ and $p_F$ so that the drift coefficient $\mu _F$, 
defined in Eq.(\ref{eq:dmudefns}), is equivalent to the ratchet velocity, $v_{BR}$:
\begin{equation}
\label{eq:TPvelocity}
\mu _F \equiv v_{BR}    
\end{equation}
Stationarity for the gap distribution is ensured by the condition $\mu _F > \mu _M$, 
which is equivalent to the condition $w_F > 0$ in the case when $\mu _M = 0$ and $p_F = 1$
(see also, Fig.\ref{fig:SepDist}(b)). In other words, the effect of the load-force and the 
consequent inward drift of the barrier particle in the BR model is mimicked in our model 
by the parameter $w_F > 0$, which also tunes the steady state value of the average gap,
$\la r \ra$ (Eq.(\ref{eq:asymptr}). So, our model can be thought of as a variant of 
the BR model \cite{peskin,mogilner} that addresses the positional fluctuations 
of the particles M and F as well.

In summary, we have discussed a two-particle model of cell protrusion namely, 
a system of two hardcore particles diffusing with different rates. 
The Brownian particle, labeled M, 
experiences a random hardcore 'push' by another particle, labeled F; 
M may be referred to as the 'driven' particle. 

We have solved the equations in the case when the drift-diffusion of particle F is small and obtained 
exact expressions for
the tagged particle moments. Physically, this corresponds to the situation when the effective 
drift of particle F towards particle M is small.  
Since particle M represents a 'membrane', fluctuations in its position is of interest when it is being 
'driven' by another particle F. Computed from the exact solutions 
(and using the ansatz Eq.(\ref{eq:sigmaMF})), our theory exhibits a crossover from an initial diffusive 
behavior to an asymptotic diffusive behavior. The existence and duration of the crossover regime 
depends on the diffusivities of the particles. It compares very well with the fluctuation data
obtained by numerically integrating a complete set of Master equations that describe the stochastic 
dynamics of this model.
Asymptotic diffusion of a tagged particle is, of course, expected for a finite system.

\section*{\leftline{\gross Appendix - Master Equations for the two-particle system}}

Let $P(n_F,n_M;t)$ denote the probability that, at a given instant of time $t$, the particles F and M are at positions $n_F$ and $n_M$ respectively. The distance between them, $n = n_M - n_F$, satisfies the inequality  $n \geq 1$ at all times; the jump probabilities per unit time, $\{ q \}$, are defined in section 2, Eq.(\ref{eq:q012}).  

\bigskip

\noindent (a) {\it Case}, $n=1$:

\begin{eqnarray}
\label{eq:neq1}
\frac{\partial P(1,2;t)}{\partial t} & = & q_1^{-}P(1,3;t) + q_0^{-}P(2,3;t) - (q_1^{+} + q_0^{+})P(1,2;t) 
                                                                                                 \nonumber\\
\\
\frac{\partial P(n_F,n_F+1;t)}{\partial t} & = & q_2^{-}P(n_F-1,n_F+2;t) + q_1^{-}P(n_F,n_F+2;t) + \nonumber \\
                                       &   & q_0^{-}P(n_F+1,n_F+2;t) + q_0^{+}P(n_F-1,n_F;t) - \nonumber \\
                                       &   & (q_2^{+} + q_1^{+} + q_0^{+} + q_0^{-})P(n_F,n_F+1;t);
                                                                          \quad (n_F\geq2) \nonumber
\end{eqnarray}

\bigskip

\noindent (b) {\it Case}, $n=2$:

\begin{eqnarray}
\label{eq:neq2}
\frac{\partial P(1,3;t)}{\partial t} & = & q_1^{-}P(1,4;t) + q_1^{+}P(1,2;t) + q_0^{-}P(2,4;t) - \nonumber \\ 
                                     &   &(q_1^{+} + q_1^{-} + q_0^{+})P(1,3;t) \nonumber \\
\\                                     
\frac{\partial P(n_F,n_F+2;t)}{\partial t} & = & q_2^{-}P(n_F-1,n_F+3;t) + q_1^{-}P(n_F,n_F+3;t) + q_1^{+}P(n_F,n_F+1;t) +\nonumber \\
                                       &   & q_0^{-}P(n_F+1,n_F+3;t) + q_0^{+}P(n_F-1,n_F+1;t) - \nonumber \\
                                       &   & (q_2^{+} + q_1^{+} + q_1^{-} + q_0^{+} + q_0^{-})P(n_F,n_F+2;t);
                                                                                        \quad (n_F\geq 2) \nonumber
\end{eqnarray}

\bigskip

\noindent (c) {\it Case}, $n\geq 3$:

\begin{eqnarray}
\label{eq:neq3}
\frac{\partial P(1,1+n;t)}{\partial t} & = & q_2^{+}P(2,n;t) + q_1^{-}P(1,2+n;t) + \nonumber \\
                                       &   & q_1^{+}P(1,n;t) + q_0^{-}P(2,2+n;t) - \nonumber \\ 
                                       &   &(q_2^{-} + q_1^{+} + q_1^{-} + q_0^{+})P(1,1+n;t) \nonumber \\
\\
\frac{\partial P(n_F,n_F+n;t)}{\partial t} & = & q_2^{+}P(n_F+1,n_F+n-1;t) + q_2^{-}P(n_F-1,n_F+n+1;t) + \nonumber \\
                                       &   & q_1^{+}P(n_F,n_F+n-1;t) + q_1^{-}P(n_F,n_F+n+1;t) +\nonumber \\
                                       &   & q_0^{+}P(n_F-1,n_F+n-1;t) + q_0^{-}P(n_F+1,n_F+n+1;t) - \nonumber \\
                                       &   & (q_2^{+} + q_2^{-} + q_1^{+} + q_1^{-} + q_0^{+} + q_0^{-})P(n_F,n_F+n;t);
                                                                                        \quad (n_F\geq 2) \nonumber
\end{eqnarray}

\bigskip

These equations could be solved for $P(n_F,n_M;t)$, subject to the initial condition $P(n_F,n_M) = \delta _{n_F,n_F^0}\delta _{n_M,n_M^0}$ at $t=0$. Since $n_F \geq 1$, we also should have $P(n_F=0,n_M;t)=0$. 

\bigskip

\noindent (d) {\it Tagged particle distributions}:

\bigskip

The tagged particle distribution functions, $P_F(n_F;t)$ and $P_M(n_M;t)$, can then be obtained from $P(n_F,n_M;t)$ using the following definitions:
\begin{eqnarray}
P_F(n_F;t) & = & \sum _{n_M = n_F+1}^{\infty} P(n_F,n_M;t) \nonumber \\
\\
P_M(n_M;t) & = & \sum _{n_F = 1}^{n_M-1} P(n_F,n_M;t) \nonumber
\label{eq:tagdist}
\end{eqnarray}

\bigskip

\noindent (e) {\it Master Equations for the separation-distance, $n$, between the particles}:

\bigskip

The probability distribution function, $P(n;t)$, for the separation distance, $n$, can be obtained from the above set of equations for the cases $n = 1, 2, 3,\cdots$ by summing over $n_F$ and also ignoring terms that violate the constraint $n_F \geq 1$. 
\begin{eqnarray}
\label{eq:sepdist}
\frac{\partial P(1;t)}{\partial t} & = & q_2^{-}P(3;t) + q_1^{-}P(2;t) - (q_2^{+} + q_1^{+})P(1;t) 
                                                                                                  \nonumber\\
\frac{\partial P(2;t)}{\partial t} & = & q_2^{-}P(4;t) + q_1^{-}P(3;t) + q_1^{+}P(1;t) - \nonumber \\
                                     &   & (q_2^{+} + q_1^{+} + q_1^{-})P(2;t) \\
\frac{\partial P(n;t)}{\partial t} & = & q_2^{-}P(n+2;t) + q_2^{+}P(n-2;t) + \nonumber \\
                                     &   & q_1^{-}P(n+1;t) + q_1^{+}P(n-1;t) - \nonumber \\
                                     &   & (q_2^{+} + q_2^{-} + q_1^{+} + q_1^{-})P(n;t); \quad (n\geq 3) \nonumber 
\end{eqnarray}
When summed up, these equations lead to the expected conservation of the total probability namely, $\sum _{n=1}^{\infty}P_n(n;t)=c$ where $c$ is a constant.

\bigskip

\noindent (f) {\it Drift and Diffusion coefficients for the case, $n\geq 3$ and $n_F \geq 2$} (second of Eq.(\ref{eq:neq3})):

\bigskip

Using the continuum variables, $x_F$ and $x_M$, for the positions of particles F and M, we can rewrite the second of Eq.(\ref{eq:neq3}) in discrete time as a jump equation,
\begin{eqnarray}
\label{eq:MEqProb}
P(x_F,x_M;N+1) & = & q_2^+ P(x_F+l,x_M-l;N) + q_2^- P(x_F-l,x_M+l;N) + \nonumber \\
               &   & q_1^+ P(x_F,x_M-l;N) + q_1^- P(x_F,x_M+l;N) + \\
               &   & q_0^+ P(x_F-l,x_M-l;N) + q_0^- P(x_F+l,x_M+l;N) \nonumber
\end{eqnarray}
where $l$ is the infinitesimal displacement of the particle.
Taylor-expanding the RHS of the above equation, and ignoring $l^3$ and higher powers, we have
\begin{eqnarray}
\label{eq:TaylorExpProb}
P(x_F,x_M;N+1) \approx P(x_F,x_M;N) & - & [w_F(2p_F-1)l] \ \frac{\partial P}{\partial x_F} + 
                                      \frac{1}{2} \ w_Fl^2 \ \frac{\partial ^2P}{\partial x_F ^2} \nonumber \\
                                    & & -[(2p_M-1)l] \ \frac{\partial P}{\partial x_M} + 
                                      \frac{1}{2}l^2 \ \frac{\partial ^2P}{\partial x_M ^2}
\end{eqnarray}
with the initial condition,
\begin{equation}
p(x_F,x_M;0) = \delta (x_F)\delta (x_M-l)
\end{equation}
For simplicity, we can take $x_F$ and $x_M$ to represent the {\it displacement} of the particles from their respective initial positions. Then the initial condition becomes
\begin{equation}
\label{eq:InCondProb}
p(x_F,x_M;0) = \delta (x_F)\delta (x_M)
\end{equation}
In the absence of the reflecting boundary at $x_F = 0$ and the hard-core constraint, the variables $x_F$ and $x_M$ are unbounded, and we define the Fourier Transform,
\begin{equation}
\label{eq:FTProb}
P(k_F,k_M;N) = \int _{-\infty}^{\infty} dx_F e^{ik_Fx_F}\int _{-\infty}^{\infty} dx_M e^{ik_Mx_M}P(x_F,x_M;N)
\end{equation}
From the initial condition, Eq.(\ref{eq:InCondProb}), it is clear that $P(k_F,k_M;0) = 1$.
\noindent Fourier-Transforming Eq.(\ref{eq:TaylorExpProb}), we get
\begin{eqnarray}
P(k_F,k_M;N+1) \approx \{ 1 & + & i[w_F(2p_F-1)l]k_F - \frac{1}{2}w_Fl^2 k_F^2 \nonumber \\
                            & + & i[(2p_M-1)l]k_M - \frac{1}{2}l^2 k_M^2\}P(k_F,k_M;N)
\end{eqnarray}
which becomes
\begin{equation}
P(k_F,k_M;N+1) \approx \left( 1 + i[w_F(2p_F-1)l]k_F - \frac{1}{2}w_Fl^2 k_F^2
                                + i[(2p_M-1)l]k_M - \frac{1}{2}l^2 k_M^2\right) ^N
\end{equation}
when iterated with respect to the discrete time variable $N$ and subject to the initial condition, Eq.(\ref{eq:InCondProb}). Written in the equivalent form, we have
\begin{equation}
\label{FTEqProb}
P(k_F,k_M;N+1) \approx \exp (N\log [1 + a_F + a_M])
\end{equation}
where
\begin{eqnarray}
a_F & \equiv & i[w_F(2p_F-1)l]k_F - \frac{1}{2}w_Fl^2 k_F^2 \\
a_M & \equiv & i[(2p_M-1)l]k_M - \frac{1}{2}l^2 k_M^2
\end{eqnarray}
Using the standard expansion for the logarithmic function,
\begin{equation}
\log (1+x) = \sum _{n=1}^{\infty} (-1)^{n+1}\frac{x^n}{n} \ ; \quad (x\ \mbox{small}) 
\end{equation}
we expand $P(k_F,k_M;N+1)$ given by Eq.(\ref{FTEqProb}):
\begin{eqnarray}
P(k_F,k_M;N+1) & \approx & \exp (N\{ a_F + a_M + \nonumber \\
                   & & \frac{1}{2}w_F^2(2p_F-1)^2l^2k_F^2 + \frac{1}{2}(2p_M-1)^2l^2k_M^2 + \nonumber \\
                   & & w_F(2p_F-1)(2p_M-1)l^2k_F k_M \})
\end{eqnarray}
where we have ignored $k_F^3,k_M^3$ and higher powers. With the following notational simplifications,
\begin{equation}
\label{coeff1}
\mu _F \equiv w_F(2p_F-1)l
\end{equation}
\begin{equation}
\label{coeff2}
\mu _M \equiv (2p_M-1)l
\end{equation}
\begin{equation}
\label{coeff3}
D_F \equiv \left(\frac{1}{2}w_F(1-w_F) + 2w_F^2p_F(1-p_F)\right)l^2 
\end{equation}
\begin{equation}
\label{coeff4}
D_M \equiv 2p_M(1-p_M)l^2
\end{equation}
we have,
\begin{equation}
P(k_F,k_M;N+1) \approx \exp (N\{i\mu _Fk_F - D_Fk_F^2 + i\mu _Mk_M - D_Mk_M^2 + \mu _F\mu _Mk_Fk_M\})
\end{equation}
which upon Fourier inversion gives
\begin{equation}
P(x_F,x_M;N+1) = g\left( \frac{x'_F}{\sqrt{4D_F N}}\right)\times 
                 g\left( \frac{x'_M - \frac{\mu _F \mu _M}{2D_F}x'_F}{\sqrt{4D'_M N}}\right)
\end{equation}
where
\begin{eqnarray}
x'_F & \equiv & x_F - \mu _F N \nonumber \\
x'_M & \equiv & x_M - \mu _M N \nonumber \\
D'_M & \equiv & D_M - \frac{\mu _F^2 \mu _M^2}{4D_F} \nonumber
\end{eqnarray}
and $g(x)$ is defined in Eq.(\ref{eq:gaussian}). It is clear that $P(x_F,x_M;N+1)$ will have a separable form only if 
$\mu _F = 0$ or $\mu _M = 0$ or both. In that case, the $\mu$'s and the $D$'s defined in Eq.(\ref{coeff1}-\ref{coeff4}) will be corresponding drift and diffusion coefficients.

%
\begin{center}
  \begin{figure}[hbt]
      \subfigure[]
                {\includegraphics[width=0.8\textwidth ]{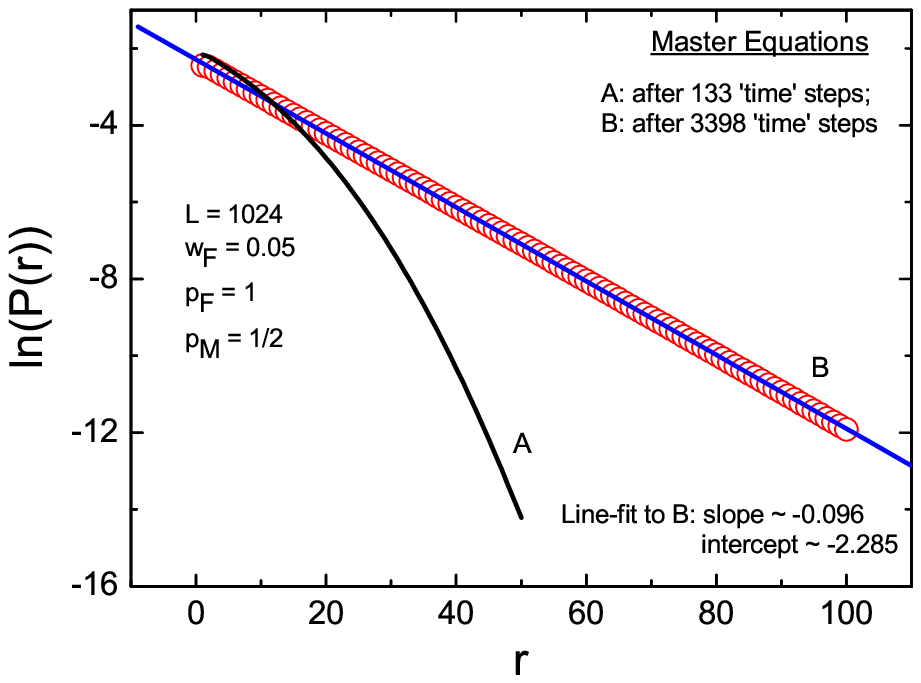}}
		
      \subfigure[]
                {\includegraphics[width=0.8\textwidth ]{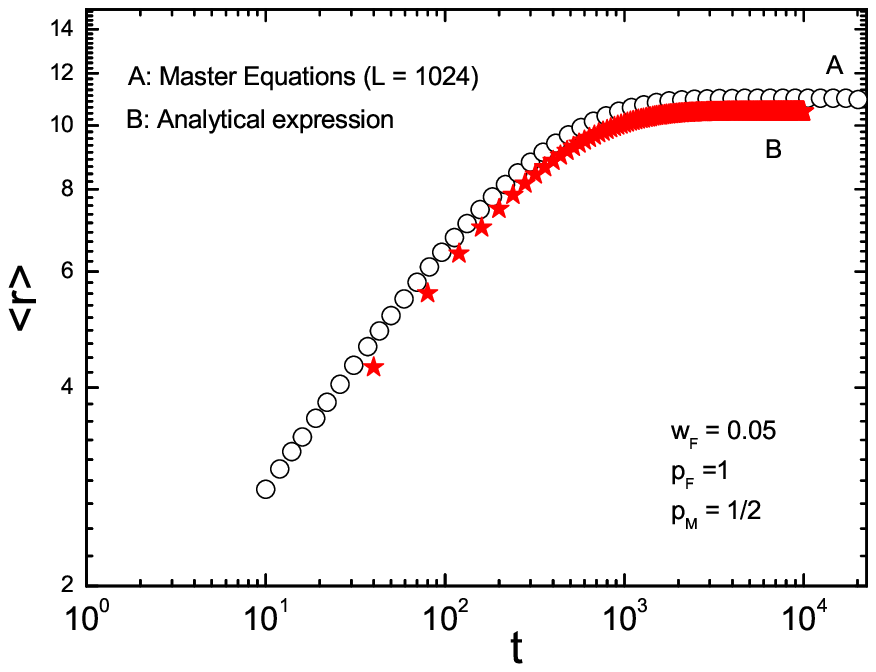}}
    \caption{(a) Probability distributions for the separation distance, $r$. 
                  Lattice size, $L = 1024$; at every   
                  time step, particle M moves left or right with equal 
		  {\it a prior} probability 1/2 whereas
                  particle F moves always to the right ($p_F = 1$) with 
		  probability $w_F = 0.05$. Open circles 
                  represent data dumped after $t = 133$ time steps ($t_1$) 
		  and also after $t = 3398$ time steps 
                  ($t_2$). Straight line fit to the data ($t_2$) indicates 
		  that $P(r) \sim 0.1e^{-0.096r}$ after
                  3398 time steps.
             (b) Average separation distance - open circles (Master equations);
	          filled stars (analytic 
                  expression). 
              The initial slope before saturation
		  is $\la r \ra \sim \sqrt{t}$.                   
            }
    \label{fig:SepDist}
  \end{figure}
\end{center}
%
%
%
\begin{center}
  \begin{figure}[hbt]
      \subfigure[]
                {\includegraphics[width=0.8\textwidth ]{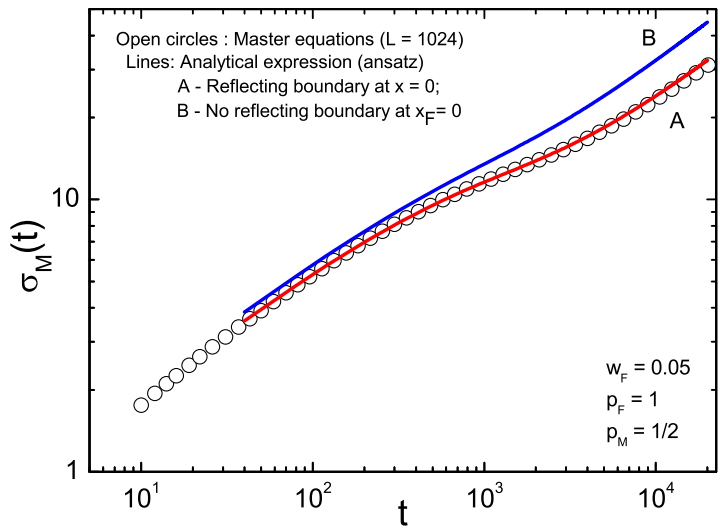}}
		
      \subfigure[]
                {\includegraphics[width=0.8\textwidth ]{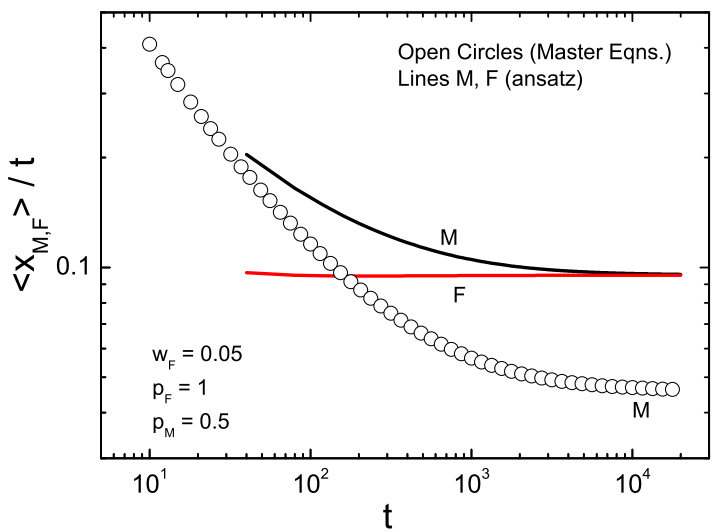}}      
    \caption{Fluctuation for the tagged particle M, $\sigma _M = (\la x_M^2 \ra - \la x_M \ra ^2)^{1/2}$. Master equations (open  
             circles); ansatz Eq.(\ref{eq:sigmaMF}) (continuous line). Parameters are the same as in Fig.\ref{fig:SepDist} - namely, 
             (a) $w_F = 0.05; p_F = 1; p_M = 1/2$. 
             (b) Average velocities of the particles, M and F, obtained by using the ansatz, Eq.(\ref{eq:AvPosMF}) (lines M and F); they  
                 approach the same value asymptotically, implying that the particles drift together as a single 
		            system. Open circles represent the average velocity of particle M obtained from the Master Equations.
                 It is clear that the asymptotic drift velocity is just proportional to that given by the ansatz. 
                 Parameters are the same as in Fig.\ref{fig:SepDist}.                                        
            }
    \label{fig:AvPosSigmaM}
  \end{figure}
\end{center}
%
%
\begin{center}
  \begin{figure}[hbt]
     \subfigure[]
        {\includegraphics[width=0.8\textwidth ]{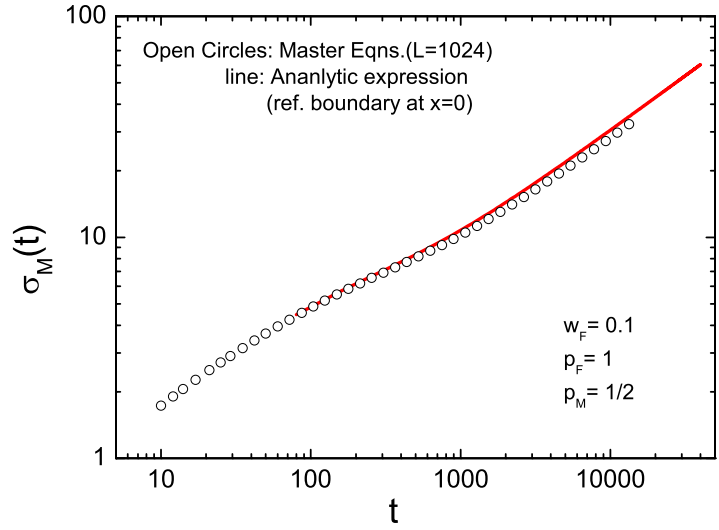}}
     \subfigure[]
        {\includegraphics[width=0.8\textwidth ]{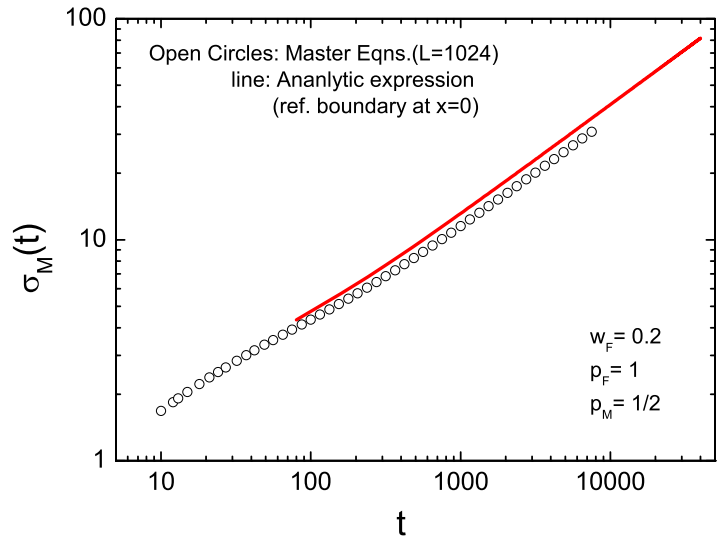}}   
      \caption{Fluctuation for the tagged particle M, $\sigma _M = (\la x_M^2 \ra - \la x_M \ra ^2)^{1/2}$. Master equations (open  
               circles); ansatz Eq.(\ref{eq:sigmaMF}) (continuous line). Parameters: (a) $w_F = 0.1; p_F = 1; p_M = 1/2$; 
                                                                                     (b) $w_F = 0.2; p_F = 1; p_M = 1/2$.}
    \label{fig:AvPosSigmaM1}
  \end{figure}
\end{center}
%
%
%
\begin{center}
  \begin{figure}[hbt]
      \subfigure[]
                {\includegraphics[width=0.8\textwidth ]{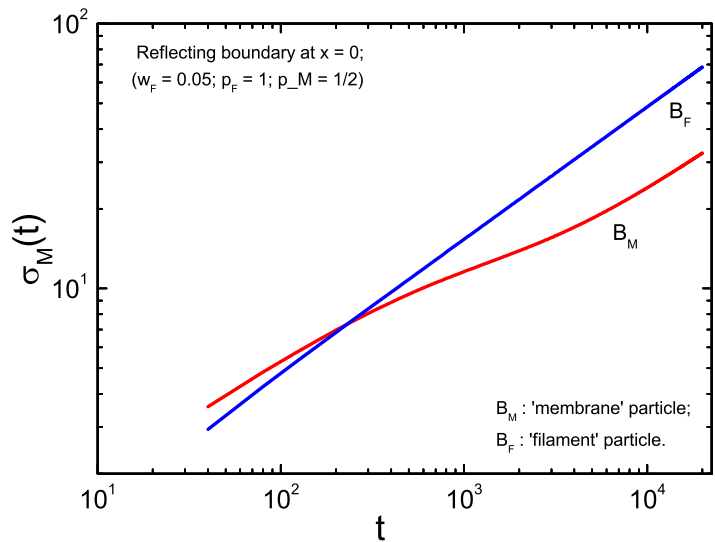}}
		
      \subfigure[]
                {\includegraphics[width=0.8\textwidth ]{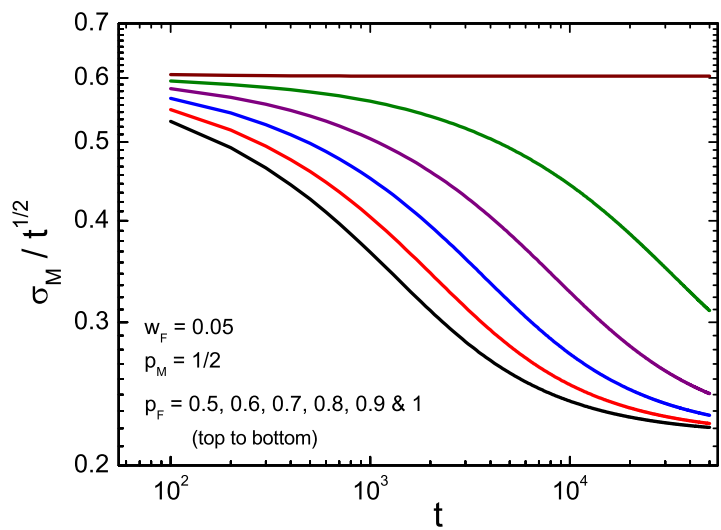}}
    \caption{(a) Computed fluctuation data (reflecting boundary at $x=0$), $B_M$ and $B_F$, for 
                  tagged particles M and F
                     respectively. It is clear that 
                     $\sigma _M \sim \sqrt{t}$ for $B_F$, whereas it 
		     shows a crossover from an initial diffusive behavior to an asymptotic ($t \sim 5000$)
                     diffusive behavior. Parameters are $w_F = 0.05; p_f = 1; p_M = 1/2$.                     
                (b) The ratio $\sigma _M/t^{1/2}$ for different values of 
		    $p_F$ ($ = 0.5, 0.6, 0.7, 0.8, 0.9, 1.0$;
                    top to bottom) with $w_F$ having a fixed 
		    value 0.05. For $p_F = 1/2$ ({\it i.e}.,
                    no effective drift towards particle M), there is no 
		    crossover. 
            }
    \label{fig:Sigma}
  \end{figure}
\end{center}
%
%
%
\begin{center}
  \begin{figure}[hbt]
    \includegraphics[width=0.8\textwidth ]{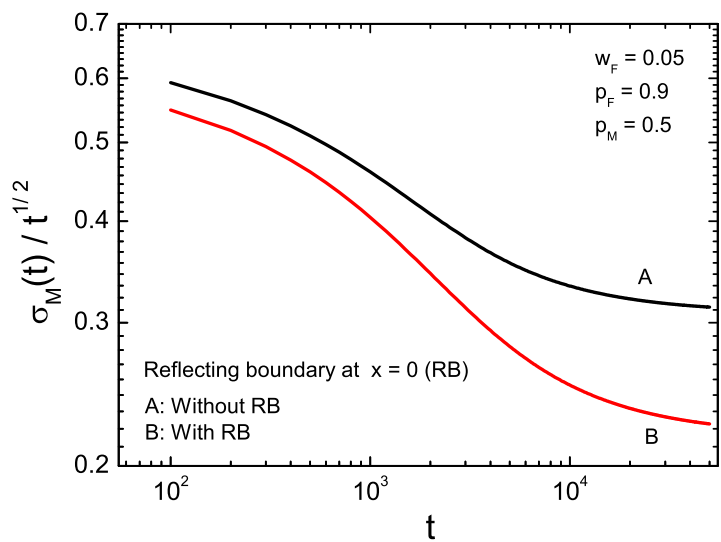}
    \caption{Fluctuation in the position of M for the cases with (B) and without (A) a 
             reflecting boundary at $x=0$. The parameters are, 
	     $w_F = 0.05; p_F = 0.9; p_M = 0.5$. The crossover is 
	     more pronounced
             for the case B than for the case A. The diffusivity of 
	     particle F, ($D_F = w_F/2$), is smaller than that of M
             ($D_M = 1/2$). 
            }
    \label{fig:SigmaRBpF09}
  \end{figure}
\end{center}
%
%

\end{document}